\documentclass[usenatbib]{mn2e}
\usepackage[dvips]{graphicx}
\usepackage[]{url}
\def\CN2{\mbox{$C_N^2 \ $}}

\title[Wind speed vertical distribution at Mt.Graham]{Wind speed vertical distribution at Mt. Graham}
\author[S. Hagelin et al.]{S. Hagelin,$^{1, 2}$ \thanks{E-mail:
   hagelin@arcetri.astro.it; masciadri@arcetri.astro.it},
 E. Masciadri$^1$\footnotemark[1], F. Lascaux$^1$ \\ $^1$INAF
 Osservatorio Astrofisico di Arcetri, Largo Enrico Fermi 5, I-501 25
 Florence, Italy\\ $^2$Uppsala Universitet, Department of Earth
 Sciences, Villav\"agen 16, S-752 36 Uppsala, Sweden}

\begin{document}
\label{firstpage}
\date{Accepted 2010 ???? ??, Received 2010 ???? ??; in original form
2010 ???? ??}
\pagerange{\pageref{firstpage}--\pageref{lastpage}}
\pubyear{2010}

\maketitle

\begin{abstract}
The characterization of the wind speed vertical distribution V(h) is
fundamental for an astronomical site for many different reasons: (1)
the wind speed shear contributes to trigger optical turbulence in the
whole troposphere, (2) a few of the astroclimatic parameters such as
the wavefront coherence time ($\tau_0$) depends directly on V(h), (3)
the equivalent velocity $V_0$, controlling the frequency at which the
adaptive optics systems have to run to work properly, depends on the
vertical distribution of the wind speed and optical turbulence. Also,
a too strong wind speed near the ground can introduce vibrations in
the telescope structures. The wind speed at a precise pressure (200
hPa) has frequently been used to retrieve indications concerning the
$\tau_0$ and the frequency limits imposed to all instrumentation based
on adaptive optics systems, but more recently it has been proved that
$V_{200}$ (wind speed at 200 hPa) alone is not sufficient to provide
exhaustive elements concerning this topic and that the vertical
distribution of the wind speed is necessary. In this paper a complete
characterization of the vertical distribution of wind speed strength
is done above Mt.Graham (Arizona, US), site of the Large Binocular
Telescope. We provide a climatological study extended over 10 years
using the operational analyses from the European Centre for
Medium-Range Weather Forecasts (ECMWF), we prove that this is
representative of the wind speed vertical distribution at Mt. Graham
with exception of the boundary layer and we prove that a mesoscale
model can provide reliable nightly estimates of V(h) above this
astronomical site from the ground up to the top of the atmosphere
($\sim$ 20 km).
\end{abstract}


\section{Introduction}
The characterization of the wind speed above an astronomical site is
extremely important for several different reasons. Firstly, because
the wind speed is strictly correlated to the optical
turbulence: strong wind speed and sharp wind speed gradients are
indicators of a turbulent atmosphere, which in combination with a
stable stratification of the atmosphere (a positive gradient of the
potential temperature) creates optical turbulence that limits the
resolution of the telescopes.  Secondly, because the stronger the wind
speed, the higher is the speed at which the turbulence layers cross
the pupil of the telescope and the higher is the frequency at which
adaptive optics systems are forced to work to correct the turbulence
perturbations above the wavefront. Thirdly, because a too strong wind speed
near the ground can introduce vibrations of the telescope structures.

At mid-latitudes, the wind speed varies with height showing a maximum
at the jet stream level, usually 10-12 km (or $\sim$200 hPa) above sea
level. The strength of the wind speed at high altitudes varies
according to season, with the strongest values during (the local)
winter and early spring \citep*{Ma01, Ca05, GL05, Ma06, Eg07, Bo09,
 Ma10}.  It has been observed that seasonal variations at the
jet stream level are more pronounced at higher latitudes than at low
latitudes in proximity of the equator \citep{Ca05} because of the
general circulation of the wind speed at synoptic scale.

At extreme latitudes, i.e. in proximity of the poles, the wind speed is
characterized by a completely different feature in the free atmosphere
\citep{Ge06, Sa06}. 
It is rather
weak up to 10 km and then increases monotonically above this height with
a rate that increases with the distance of the site from the centre of
the polar vortex \citep{Ha08}. 
This assumption has been confirmed by \cite{La09}. 

Apart from the fact that a windy atmosphere more easily triggers optical
turbulence, the astroclimatic parameter that directly depends not only
on the optical turbulence but also on the wind speed is the wavefront
coherence time:
\begin{equation} \label{tau}
\tau_0 = 0.057 \lambda^{6/5} \Big(\int\limits_{0}^{\infty }{} V(h)^{5/3} C_N^2(h) dh \Big)^{-3/5}
\end{equation}
also equivalent to:
\begin{equation}
\tau_{0}=0.31\frac{r_{0}}{V_{0}}
\end{equation}
where $r_{0}$ is the Fried's parameter and V$_{0}$ is the equivalent velocity:
\begin{equation}
V_{0}=\left( \frac{\int\limits_{0}^{\infty }{}V(h)^{5/3}
C_{N}^{2}\left( h\right) dh}{\int\limits_{0}^{\infty }C_{N}^{2}dh}\right)
^{3/5}
\label{vel}
\end{equation}

$\tau_{0}$ depends, therefore, on the vertical distribution of the
wind speed V(h) and the optical turbulence $\CN2$(h). To simplify the
calculation of the wavefront coherence time Sarazin \& Tokovinin
(2002) proposed a method to calculate $\tau_{0}$ using only the wind
speed estimated at 200 hPa instead of the whole profile using an
empirical relationship between $V_{0}$ and $V_{200}$: firstly they
used $V_0=0.4*V_{200}$ and, in a second time, the expression has been
modified in $V_0=max(V_{ground},0.4*V_{200})$.  However, a few years ago
\citep{Ma06}, in a study done above San Pedro M\'artir, it has been
observed that the value of the constant 0.4 was not universal.  The authors found 
the value 0.56 above San Pedro M\'artir. However, they also proved that the relative error introduced in $\tau_{0}$
with the method of \cite{Sa02}, that does not consider the vertical distribution of the wind speed but
just the wind speed in two precise regions of the atmosphere (near the ground and at 200 hPa), could be as great
as 20$\%$-50$\%$, even using the appropriated constant. In other words, they proved that the proportionality between V$_{0}$ and V$_{200}$ is poorly reliable if one wishes precise 
estimates of  $\tau_{0}$ (even if we select an appropriated constant)\footnote{More recently, other authors \citep{GL09} calculated
$V_0=max(V_{ground},0.4*V_{200})$ 
above the Teide Observatory and they found a third different value of the constant. However, we think they misunderstood the \cite{Ma06} thesis because they stated that the \cite{Sa02} method needs to be calibrated to be used.}.
\cite{Ma06} concluded therefore that, for the calculation of the $\tau_{0}$, the
vertical distribution of the wind speed on the whole troposphere
is fundamental and necessary and the method suggested by 
\cite{Sa02}  can provide only some qualitative (but not accurate) estimates because it presents some intrinsic weak
points. It appears therefore absolutely important in
the field of the site characterization of an astronomical site, at least in cases such as the calculation of $\tau_{0}$, to
characterize the wind speed vertical distribution. How to do that?

The estimate of the wind speed vertical distribution up to 20 km at an
astronomical site (usually placed on the top of mountains) is not
trivial.  The Generalized Scidar, an optical instrument based on a
remote sensing principle, can measure the wind speed \citep*{Av01}
at all the heights in which turbulent layers are present i.e. it can 
reconstruct  a sort of vertical profile. Some estimates have been done
in the past above different sites \citep{Av06, Eg07, Ma10}.
However, this method requires the employment of an instrument that has
to be placed at the focus of a telescope with a pupil size of at least
1.5 m and it can be performed for short periods (typically some tens
of nights) related to dedicated experiments. Such an instrument is not
suitable for routinely monitoring the wind speed or climatological
studies. Alternatively wind speed profiles are routinely calculated by
the General Circulation Models (GCM) (mainly from the ECMWF and
NOAA/NCEP) and data-set can be retrieved in any site in the world with
a horizontal resolution of 0.25 degrees (operational analyses) and 2.5
degrees (re-analyses)\footnote{To obtain re-analyses the GCM models
  are re-run as to offer an equal horizontal and vertical resolution
  and the same model configuration for many years in the past. In this
  way, the GCM outputs referring to the most recent years and the GCM
  outputs referring to older periods are made uniform. In other words,
  the performances of the re-analyses are degraded with respect to
  operational analyses to preserve the temporal uniformity. On the
  other hand the data-set is uniform all along decades and this is the
  reason why re-analyses are used mainly for climatological
  studies.}. Both have been used for astronomical applications. More
precisely, the ECMWF analyses have been used at mid-latitude sites
\citep{Ma01, Eg07, Ma10}
as well as at extreme latitudes \citep{Ge06, Ha08}.
NOAA/NCEP re-analyses (vertical profiles) have been used at
mid-latitude sites \citep{Ca05, Av06, Bo09} and ERA-reanalyses of the ECMWF 
have been used at extreme latitudes \citep{Sa06}.
Model outputs showed good correlations with measurements in all cases
in which this has been calculated. The unique problem with analyses
and re-analyses is that these estimates are representative of the wind
speed above the astronomical site but not in the the surface and
boundary layer where the local orographic effects have a major effect
on the wind speed. Above roughly 1-2 km from the ground, the wind
speed is almost horizontally homogeneous and the vertical distribution
is basically the same with respect to a horizontal extension of some
tens of kilometers. Below this height the analyses from the GCMs are
less representative of the wind speed because their horizontal
resolution is too low to give an accurate description of the
interaction of the atmospheric flow with the topography near the
surface.

However, in the surface (lowest few tens of meters) and boundary layers (typically the first kilometer above the surface) it has been proved \citep{Ma03} 
that mesoscale models, with a horizontal resolution of 1 km, can
provide much better estimates than what analyses from the GCMs do
above mid-latitude astronomical sites. In that paper the
author proved that mesoscale models can provide estimates better
correlated to measurements than analyses from GCM. It has also been
proved that mesoscale models are able, contrary to the analyses
from the GCMs, to discriminate the wind speed near the ground between
two astronomical sites (Paranal and Maidanak) characterized by a
median wind speeds that differ for 4-5 ms$^{-1}$. This study lets us think
that mesoscale models could be a useful tool to reconstruct the wind
speed vertical profile all along the 20 km from the ground.  They are
supposed to be comparable in performances to the ECMWF models above
1-2 km from the ground and they are supposed to provide better
performances of the wind speed in the first 1-2 km from the ground.

In this paper we provide a climatological characterization
of the wind speed above 1 km from the ground on the time scale of ten years using analyses from the ECMWF general circulation models and
we investigate the possibility to use a mesoscale model to
systematically reconstruct a complete wind speed vertical profile
extended on the whole 20 km above Mt.Graham (Arizona, US) the site of
the Large Binocular Telescope (LBT).

In Section 2 we use the operational ECMWF analyses over 10 years,
1998-2007, to present a climatological estimate of the median monthly wind
speed. The operational data was chosen because of their higher
resolution (0.25 degrees) with respect to the re-analyses. We
will first prove the reliability of the ECMWF analyses in these
regions of the world comparing analyses with radiosoundings launched in
from Tucson International Airport ( $\sim$120 km from Mt. Graham).

In Section 3 we investigate the reliability of the wind speed vertical
profiles retrieved from simulations with a mesoscale model (Meso-NH)
in the high as well as in the low part of the atmosphere. The
mesoscale model is run in a grid-nesting configuration covering a
total surface of 800 km x 800 km and three imbricated models with 10,
2.5 and 0.5 km horizontal resolution (Table \ref{tab1}). The innermost
model covering a surface of 60 km x 60 km.  To estimate the model
reliability we use measurements of the wind speed done by a
Generalized Scidar run at the focus of the Vatican Advanced Technology
Telescope (VATT) as well as by an anemometer located on the roof of
the same telescope.  This section aims to evaluate the possibility to
use a mesoscale model to characterize the vertical distribution of the
wind speed extending from the ground up to 20 km above astronomical
sites. This should certainly represent an extremely valuable tool to
provide an exhaustive monitoring of the wind speed above an
astronomical site.

\begin{table}
\caption{Meso-NH model configuration. In the second column the  horizontal resolution $\Delta$X, in the third column the number of grid points and in the fourth column the horizontal surface covered by the model domain.}
\begin{tabular}{cccc}
\hline
& $\Delta$X (km) & Grid Points & Surface (km)\\
\hline
model 1 & 10  & 80 x 80 & 800 x 800\\
model 2 & 2.5 & 64 x 64 & 160 x 160 \\
model 3 & 0.5 & 120 x 120 & 60 x 60\\
\hline
\end{tabular}
\label{tab1}
\end{table}

Finally, in Section 4, we present the conclusion of this study.


\section{Monthly median wind speed}
\label{ses2}

The ECMWF analyses used in this study are the operational analyses,
downloaded from the MARS archive.\footnote{http://www.ecmwf.int} For a
more detailed description of the data-set we refer the reader to 
\citep*{Ma01, Ge06}. As explained in the Introduction we only consider
the wind speed above 1 km from the ground in this section.

In a previous study \citep{Eg07} done at Mt. Graham, it has been proved that independent measurements
of the wind speed done at the summit of the mountain with a
Generalized Scidar have a good and small relative discrepancy of the order of
23$\%$ with respect to the ECMWF analyses extracted from the nearest
grid point.  Similar results have been obtained above San Pedro
M\'artir by \cite{Av06}. 

\begin{figure*}
\includegraphics[width=7cm]{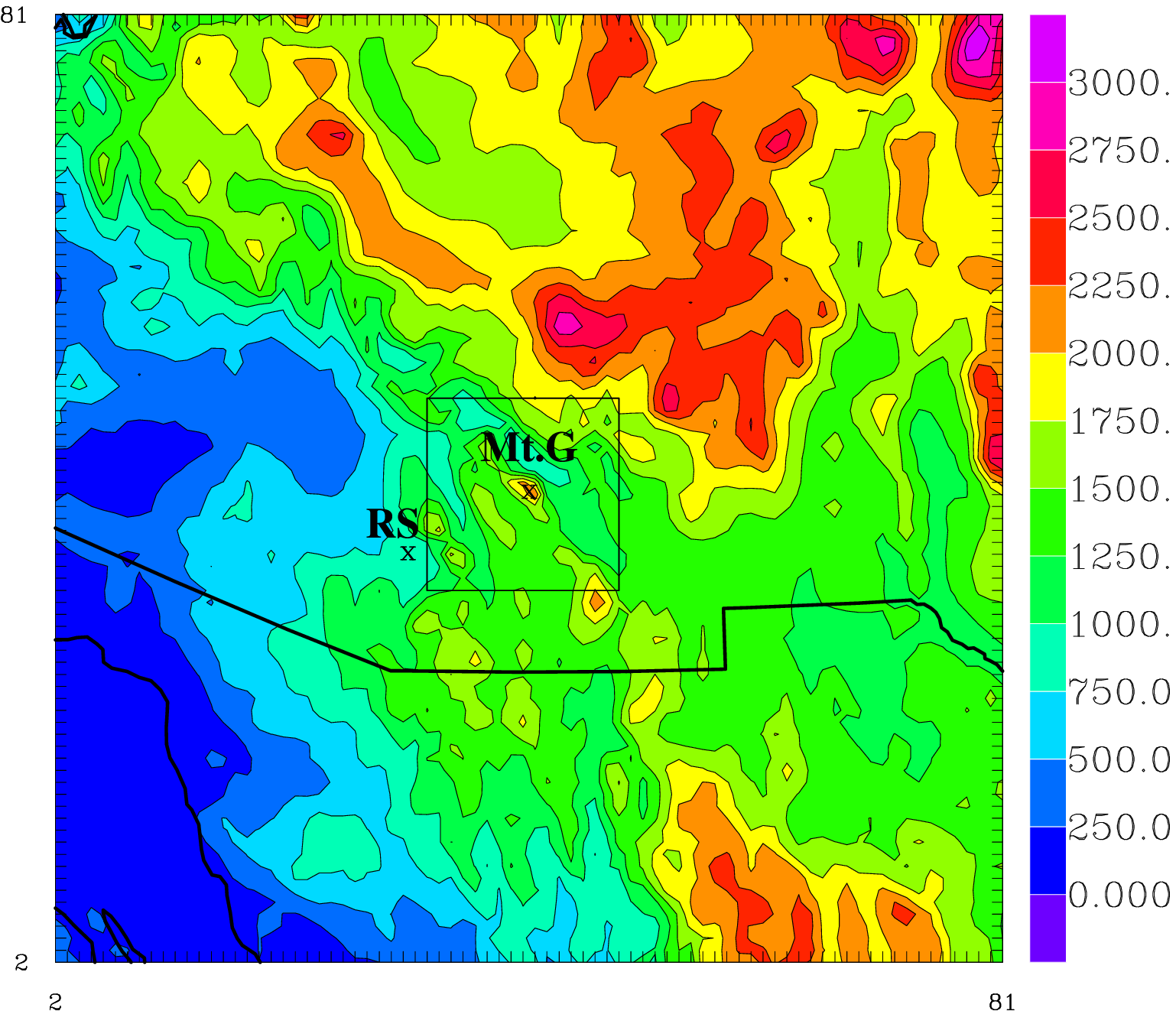}
\includegraphics[width=7cm]{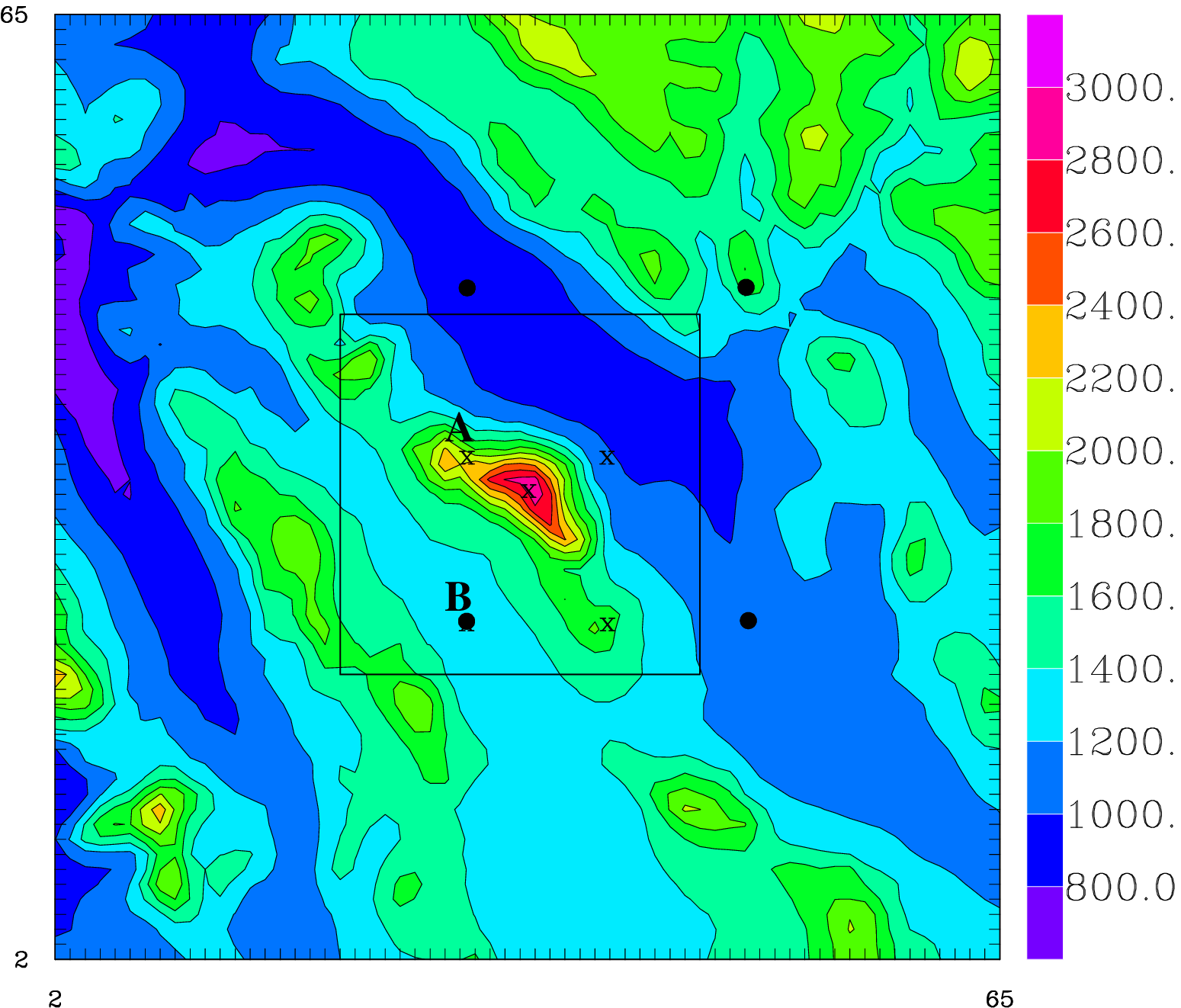}
\includegraphics[width=7cm]{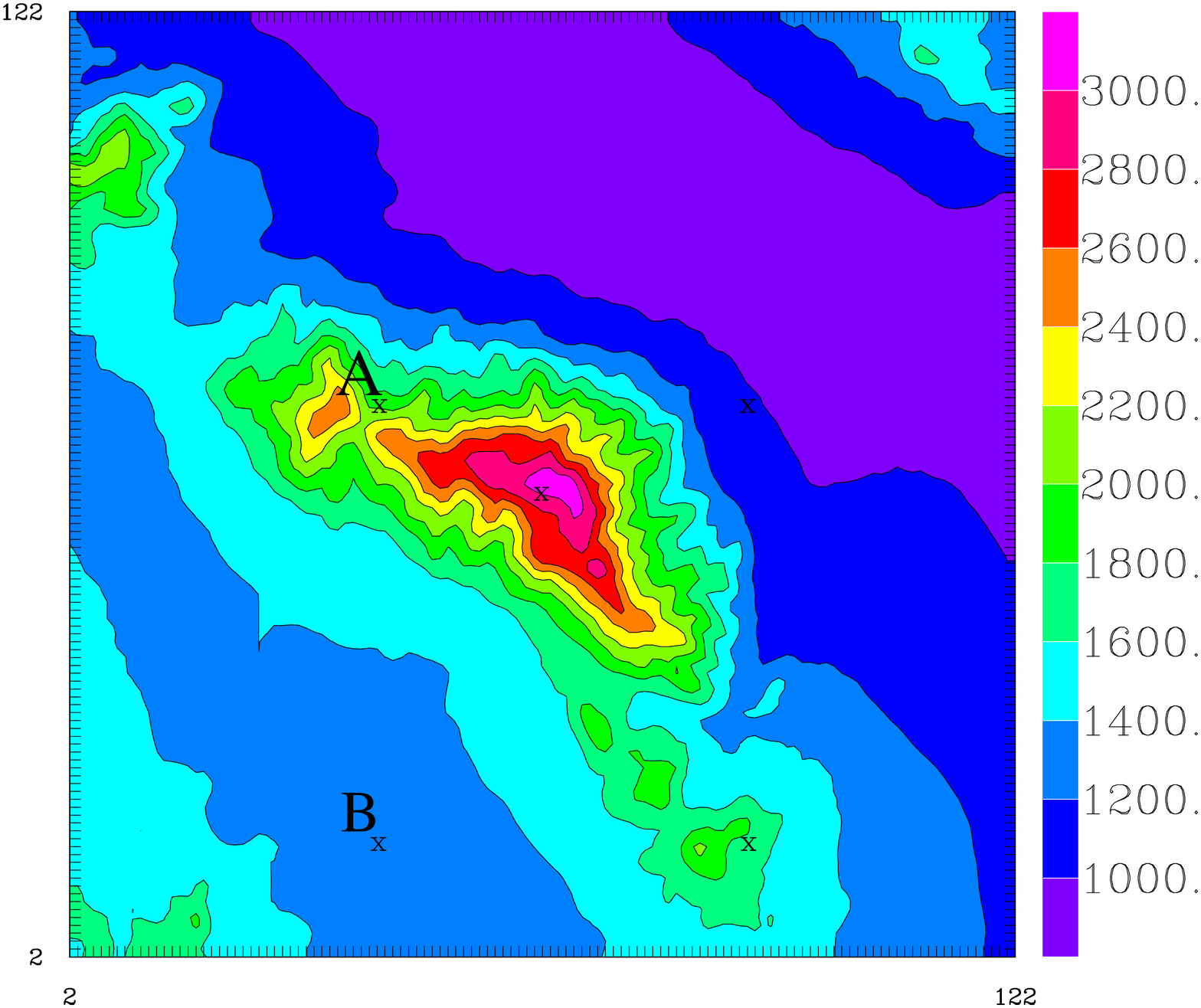}
\caption{{\bf Top Left:} Digital elevation map (DEM) of the Mt.Graham
  region (model 1 of Table \ref{tab1} extended on 800 km x 800 km with
  $\Delta$X=10 km).  The distance between the Mt. Graham summit and
  the Tucson Airport Station (label RS) is $\sim$ 120 km. The black
  square shows the location of model 2.  {\bf Top Right:} DEM (model 2
  of Table \ref{tab1} extended on 160 km x 160 km with $\Delta$X=2.5
  km).The black square shows the location of the innermost model
  (model 3 of Table \ref{tab1} extended on 60 km x 60 km with
  $\Delta$X=0.5 km). The location of the Mt.Graham Observatory is
  marked with an {\it x} on the summit of the mountain in the middle
  of the figure. The grid-points in which ECMWF analyses have been
  extracted (when the GCM resolution is 0.25$\degr$) are also marked
  with a {\it x}. The grid-points in which ECMWF analyses are
  extracted (when the GCM resolution is 0.5$\degr$) are marked with
  black dots.  {\bf Bottom:} DEM (model 3 of Table \ref{tab1} extended
  on 60 km x 60 km with $\Delta$X=0.5 km).}
\label{map}
\end{figure*}

\begin{figure*}
\includegraphics[width=7cm, angle=-90]{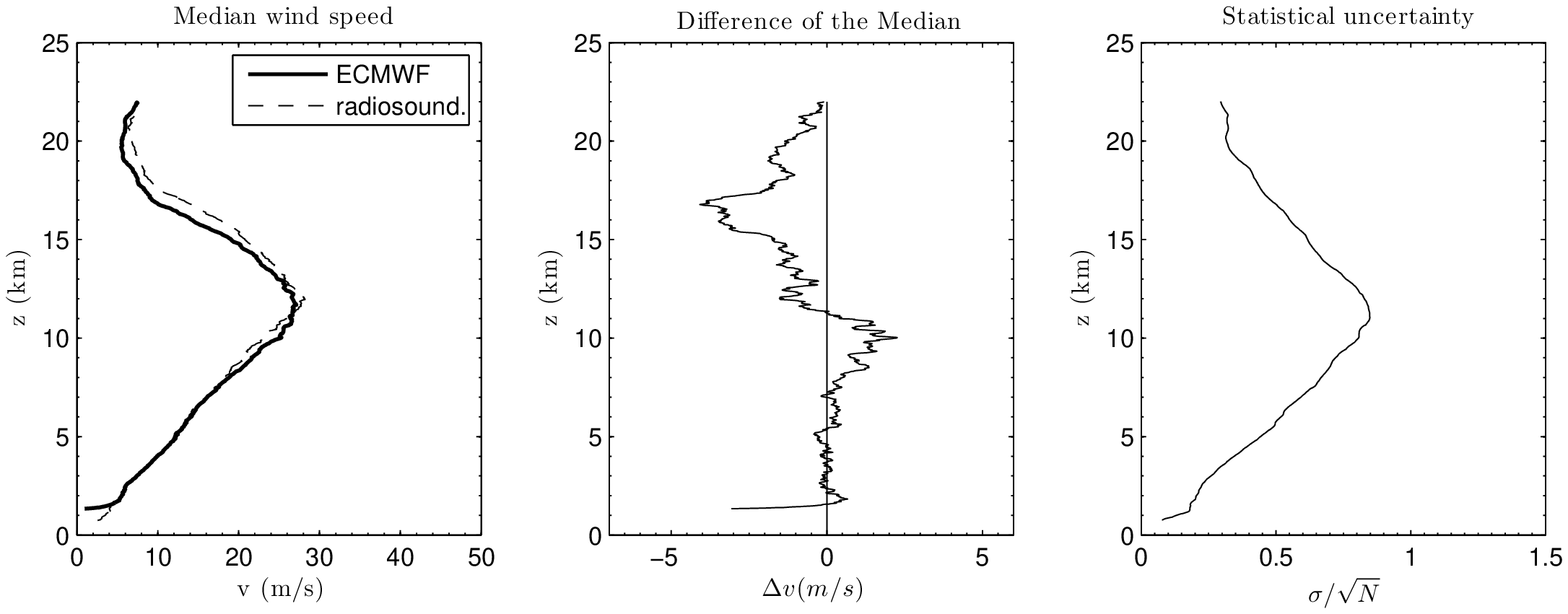}
\caption{{\bf Left:} median wind speed extended on the whole year 2005
  calculated for the radiosoundings and the ECMWF analyses (see
  text). {\bf Centre:} difference of the median wind speed (radiosoundings
  minus ECMWF analyses). {\bf Right:} statistical uncertainty calculated for
  N=328, where N is the number of the nights for which we consider
  radiosoundings.}
\label{rs1}
\end{figure*}

We try here a deeper analysis. The quality of the analyses in a region
of the earth depends on how far/close are the meteorological stations
with respect to the site one is studying and on the density of the
measurements, i.e the number of the meteorological station in that
particular region. In other words, it depends on the observations that
have the most important weight in the data assimilation process\footnote{The
  data assimilation is the procedure that, in a forecast model,
  prevents it from drifting away from the true state of the
  atmosphere} of the GCMs for that region. The nearer to the
astronomical site are the meteorological stations from which the
radiosoundings are launched, the better is the quality of the analyses
in proximity to the astronomical site. In order to make an estimate of
how well the ECMWF analyses perform in the region of Mt.Graham we have
made a comparison between the radiosoundings of the closest available
meteorological station: Tucson airport (32.23$\degr$N, 110.96$\degr$W)
at $\sim$120 km southwest from Mt.Graham (32.70$\degr$N,
109.89$\degr$W) (Fig.\ref{map}-top left panel) and ECMWF analyses
extracted from the closest grid point to the peak of Mt. Graham
i.e. (32.75$\degr$N, 110.00$\degr$W) at $\sim$ 12 km north-west from
the Mt. Graham peak (Fig.\ref{map}-right panel, point with label
(A)). The radiosoundings are available at 00:00 and 12:00 {\sc
  utc}. We considered those at 12:00 {\sc utc} corresponding to night
time conditions at local time (12:00 {\sc utc} equals 05:00 {\sc
  mst}).  Fig.\ref{rs1} shows the median wind speed vertical profile
related to the whole year 2005 (left-panel), the difference of the
median values (central-panel) and the statistical uncertainty
$\sigma$/$\sqrt N$ that measures the accuracy of the measurements
(right-panel).  We observe that the difference of the median value is
always very small (of the order of 1-2 ms$^{-1}$) in most part of the
20 km with a relative discrepancy less than or equal to 13$\%$.  Only at
17 km the difference of the median values is somehow larger ($\sim$ 4
ms$^{-1}$). However, as can been seen in Fig.\ref{rs1}-left panel, at
this height the wind speed strength is almost 1/3 of the value assumed
at the jet-stream level and it affects the integrated astroclimatic
parameters in a less important way.  In the calculation of the
statistical uncertainty $\sigma / \sqrt{N}$, N $=$ 328 is the number
of nights with radiosoundings available during 2005. We eliminated 37
nights for which the radiosoundings did not cover the whole 20 km
above sea level. We conclude, therefore, that, above 1 km from the
ground, the ECMWF analyses provide an accurate estimate of the wind
speed and the wind speed vertical distribution is uniform on a
horizontal scale of a few tens of kilometers.

\begin{figure*}
\includegraphics[angle=-90,width=175mm]{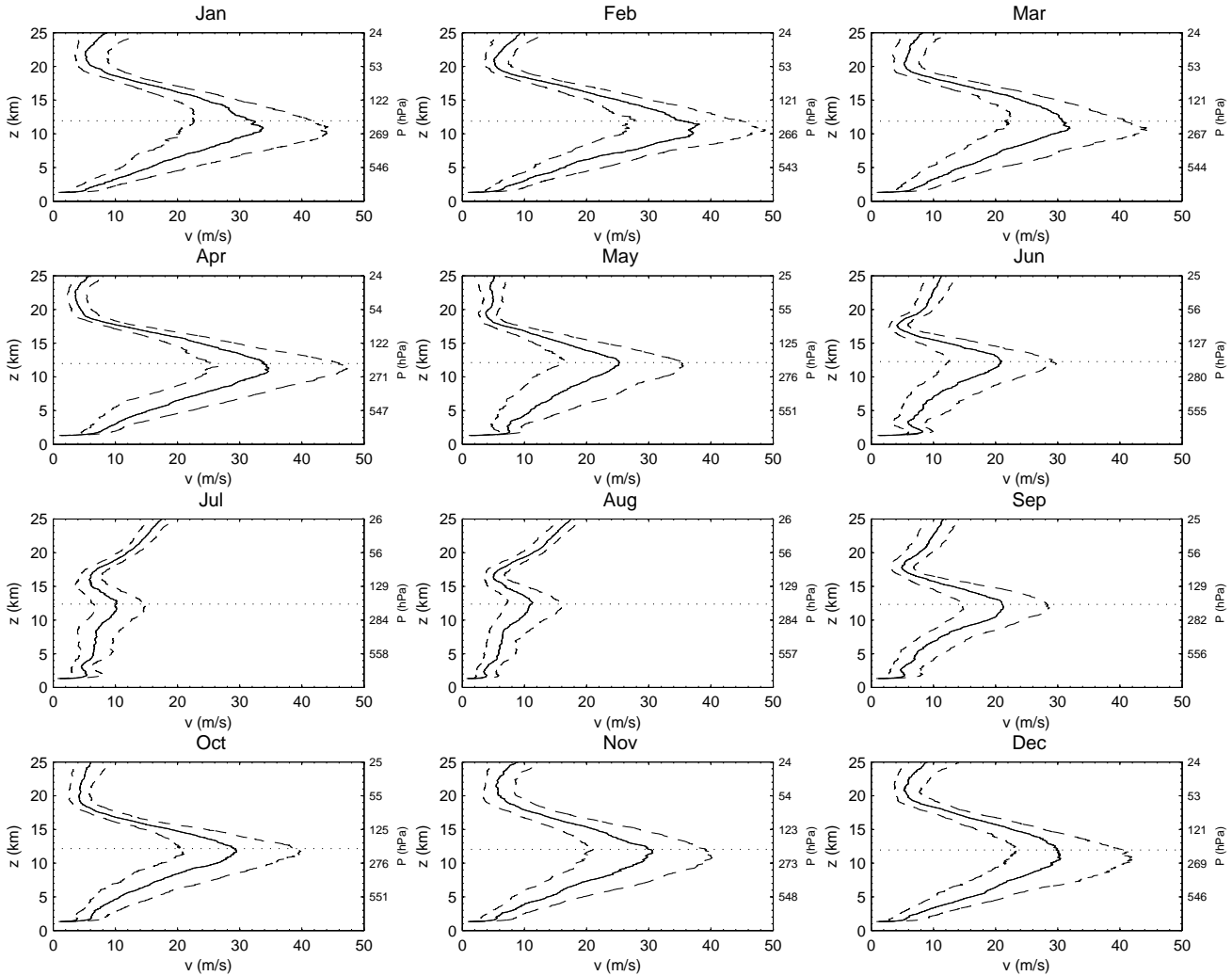}
\caption{The median wind speed calculated with the ECMWF analyses for each month during 1998-2007 at Mt
 Graham, solid line. The dashed lines represent the first and third
 quartile. The dotted line indicates 200 hPa. Data-set are extracted from the grid point indicated with the label (A) in Fig.\ref{map}-right panel.
\label{wind10}
}
\end{figure*}

Fig.~\ref{wind10} shows the monthly median vertical wind speed profile
from the ECMWF-analyses for every month calculated on a time scale of ten years, from 1998 to
2007 at 06 {\sc utc} (23 {\sc mst}), from the ground up to 25
km. Considering the difference in altitude between the ECMWF
grid-point ($\sim$ 1320 m) and the Mt. Graham summit (3200 m) we
discuss these results for h $>$ 4 km above the sea level i.e. roughly
1 km above the ground of the highest location (Mt. Graham).  The wind speed
profiles calculated above Mt. Graham are characteristic of a
mid-latitude site and the maximum of the wind speed is observed at 11-12 km above
sea level. The wind
speed maximum is located at the height of the jet-stream for the most
part of the year. During July and August the wind speed at the
jet-stream level is much weaker than during the other months and, in this period, the
highest wind speed value is observed above 20 km, well into the
stratosphere. The wind speed at the jet-stream level follows the
classical sinusoidal seasonal trend in different periods of the year.
Table~\ref{tab_200} reports, for each month of the year, the average
of the wind speed, the standard deviation for the yearly average
values and the height corresponding to 200 hPa.

The wind speed at this height is characterized by values comparable to
those observed above the major astronomical sites \citep{Ca05}. 
The month with the strongest wind speed at 200 hPa is February (37.21
ms$^{-1}$), while the weakest wind speed is observed in July (11.17
ms$^{-1}$). The mean wind speed
values as well as the standard deviation in each month
appear very similar to that observed at San Pedro M\'artir\footnote{We precise, for
 correctness, that studies in  \cite{Ca05}
are done with re-analyses while our study is done with operational analyses and
that we do not study exactly the same time period.}.
This is not surprising considering that the two sites are located
close to each other.

\begin{table}
\caption{The mean wind speed at 200 hPa (200 mbar) at Mt Graham for the period 1998-2007. 
Column three and four are the standard deviation of the wind speed and the height corresponding to 200 hPa.} 
\begin{center}
\begin{tabular}{cccc}
\hline 
Month & Avg (ms$^{-1}$) & std (ms$^{-1}$) & h (km)\\
\hline
January   & 32.79 & 4.22  & 11.92 \\ 
February  & 37.21 & 4.55  & 11.88 \\
Mars      & 32.65 & 6.70 & 11.89 \\
April     & 36.09 &  6.81& 11.98 \\
May       & 26.77 &6.54  & 12.11 \\
June      & 21.64 & 5.36 & 12.26 \\
July      & 11.17 &  2.08 & 12.37 \\
August    & 11.80 &2.82  & 12.37 \\
September & 22.77 &  3.42 & 12.31 \\
October   & 29.79 & 5.85 & 12.16 \\
November  & 30.88 &  6.25& 12.04 \\
December  & 32.21 & 3.20 & 11.93 \\
\hline
Average   & 27.06 & 4.82 & 12.10 \\
\hline
\end{tabular}
\end{center}
\label{tab_200}
\end{table}

\begin{figure*}
\includegraphics[angle=-90,width=175mm]{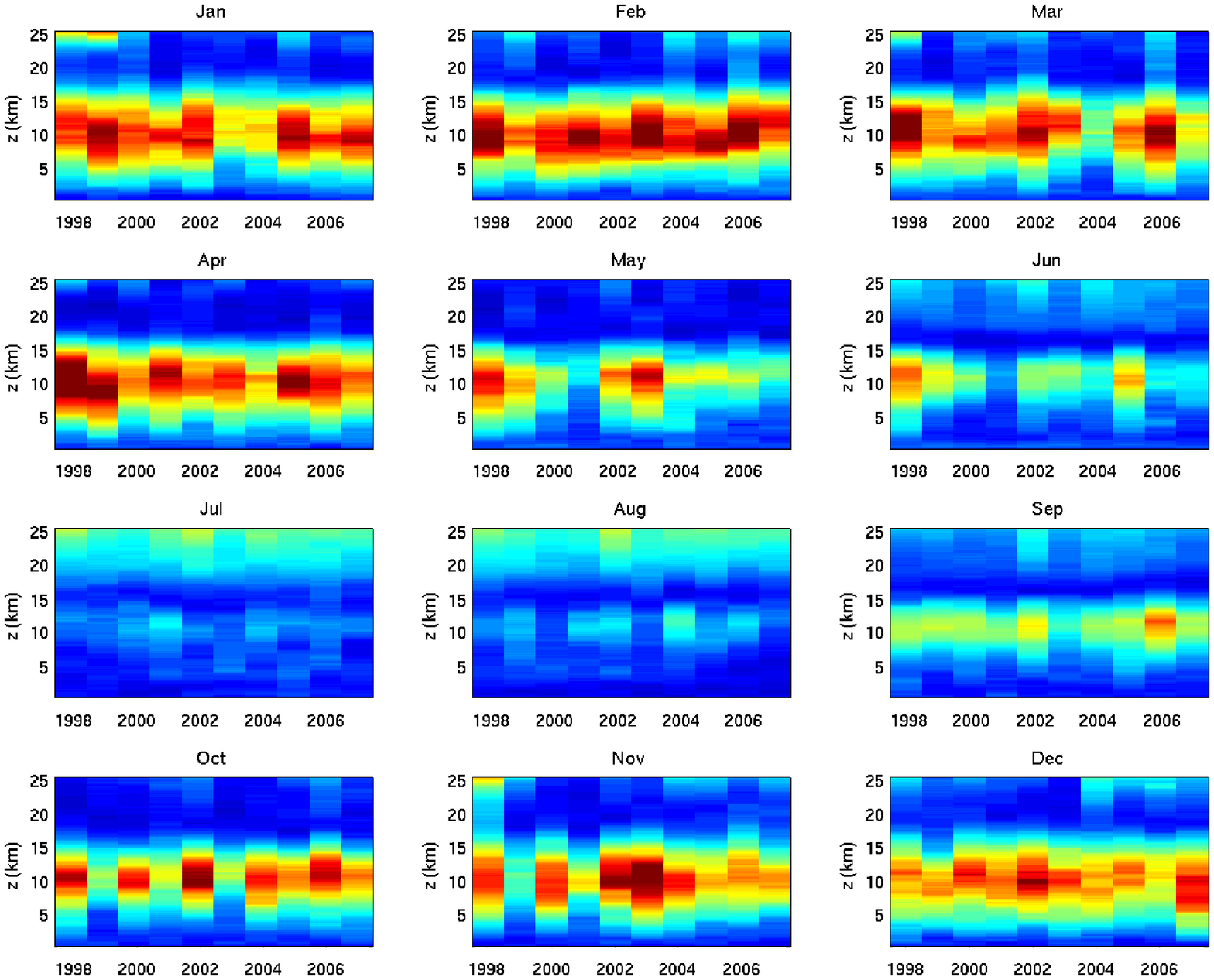} 
\includegraphics[width=130mm]{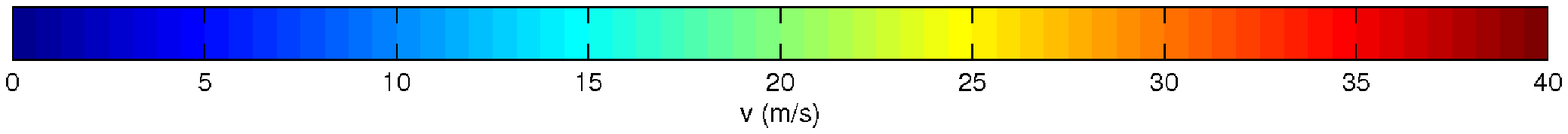}
\caption{The temporal evolution of the annual median wind speed on the 1998-2007 period calculated 
with ECMWF analyses at Mt
 Graham. The color scale goes from 0 ms$^{-1}$ (blue) to 40 ms$^{-1}$ (dark
 red). 
\label{wind_y}
}
\end{figure*}

The monthly median wind speed profile for each year is shown in
Fig.~\ref{wind_y}. The difference between the individual years is
smallest during the summer, when the strength of the wind speed is
also the weakest, as is also indicated by the smaller difference
between the first and the third quartiles observed for these months in
Fig.~\ref{wind10}. The greatest difference between individual years
occurs during autumn (October and November) and spring (March and
April). The difference between the calmest and the strongest yearly
wind speed at the jet-stream level is over 20 ms$^{-1}$ during these
months.  This is coherent with the dispersion (dashed lines) indicated
in Fig.~\ref{wind10}. A peculiar pattern of alternating years with
stronger and weaker winds is also found in October. The other months
do not show this pattern, but there are clearly rather large
differences in-between different years. This result indicates that
studies on seasonal trends, for the wind speed and also the optical
turbulence (that depends on the wind speed), should be preferably done
on time scale of the order of some years to filter out effects due to
this intrinsic variability that characterizes the wind speed in each
month.

It is worth noting that the horizontal resolution of ECMWF-analyses,
has changed during the studied period
(1998-2007)\footnote{\url{http://www.ecmwf.int/products/data/technical/model_id/index.html}}.
During this period the horizontal resolution changed from 0.5 $\degr$
to 0.25$\degr$.  Figure \ref{map} shows the locations of the four grid
points that are closest to Mt.Graham in the case of a resolution of
0.25$\degr$ (crosses) and 0.5$\degr$ (black points). With a horizontal
resolution of 0.25$\degr$, the closest grid point to the peak of the
mountain is A=(32.75$\degr$N, 110.00$\degr$W), located west-northwest
of the peak. With a horizontal resolution of 0.5$\degr$ the closest
grid point is B= (32.50$\degr$N, 110.00$\degr$W), southwest of the
peak. A comparison of the vertical wind speed profile from both these
grid points (see Appendix A) show that they are almost identical in
the free atmosphere, therefore we consider that there are no problems
of data inhomogeneity when we treat the average on time scale of the order of ten years.

\section{The wind speed from the Meso-NH model}

The Meso-NH is a non-hydrostatic mesoscale model developed jointly by
M\'et\'eo-France and Laboratoire d'A\'erologie \citep{La98}.  It is a
grid point model based on the anelastic approximation that can
simulate the temporal evolution in three dimensions of the classic
meteorological parameters such as wind speed and direction, potential
temperature and pressure.

For this study we have run the Meso-NH model in grid-nesting mode,
using three two-way nested models, centered at the peak of Mt
Graham. Details of the horizontal size and resolution of the three
models are reported in Table ~\ref{tab1}. The outermost model covers
an area of 800 x 800 km. The areas covered by the three models are
shown in Fig.~\ref{map}. The vertical grid is composed of 49 levels
covering up to 20 km above sea level.  The first vertical grid point
is located at 20 m. Above we have a logarithmic stretching of 20\% for
the vertical grid size up to 3500 m\footnote{The logarithmic stretching creates a vertical grid where the distance between each level is 20\% more than between the previous two.
}. Above 3500 m the distance
between the vertical levels is fixed and equal to 600 m.

The model is initialized with the analyses of the ECMWF, which also
provides the boundary conditions for the outermost model. The runs
start at 00:00 {\sc utc} (17:00 {\sc mst}), the synoptic hour closest
to local evening, and last for 12 hours, to early morning in Arizona
(05:00 {\sc mst}). The first two hours are rejected to avoid the
calculations being affected by spurious values due to the adaptation
of the atmospheric flow to the ground. Data from the innermost model
with the highest horizontal resolution (model 3) are treated here.

\begin{figure*}
\includegraphics[width=13cm]{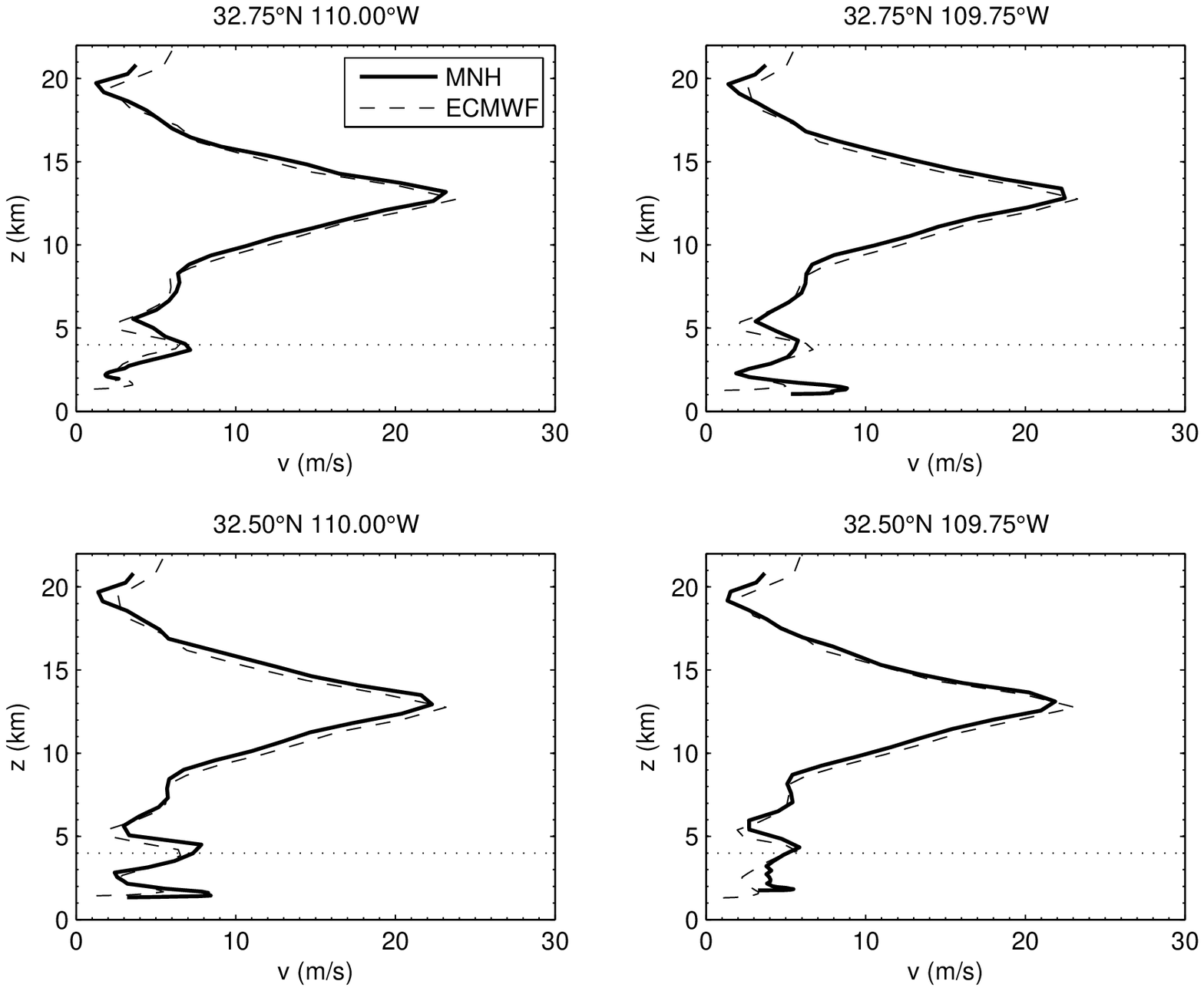}
\caption{The vertical wind speed profile for the four grid points from
  the Meso-NH (solid line) and the ECMWF analyses (dashed line) that
  are located closest to Mt.Graham (grid points indicated with a cross
  in Fig.\ref{map}), when using the 0.25\degr resolution. Calculations
  are done for the night May 21, 2005. The dotted line indicate the
  typical height below which the comparison makes no sense.}
\label{wind_4gp}
\end{figure*}

To ensure that the Meso-NH model can reconstruct reliable wind speed
profile for h $\ge$ 1 km we compare the wind speed calculated by
Meso-Nh with the wind speed calculated by the ECMWF analyses in the
four grid points that surround Mt.Graham using the 0.25$\degr$
resolution (the four grid points are indicated with a {\it x} in
Fig.~\ref{map}). Figure~\ref{wind_4gp} shows this comparison
calculated for one night, May 21, 2005. The Meso-NH profiles are
obtained by averaging the outputs calculated at each hour for the
entire 10 hour simulation covering the period (17:00 to 05:00 {\sc
  mst}). The ECMWF data are only available at the synoptic hours,
therefore we considered the average of the ECMWF data at 00, 06 and 12
{\sc utc} (at 17, 23 and 05 {\sc mst}).

We observe that the wind speed profile from Meso-NH
is very well correlated with the ECMWF analyses for all the four grid
points above roughly 1 km from the ground. This also indicates that 
the wind speed is horizontally homogeneous over a surface of 0.25$\degr\times$0.25$\degr$.

\begin{figure*}
\includegraphics[width=13cm]{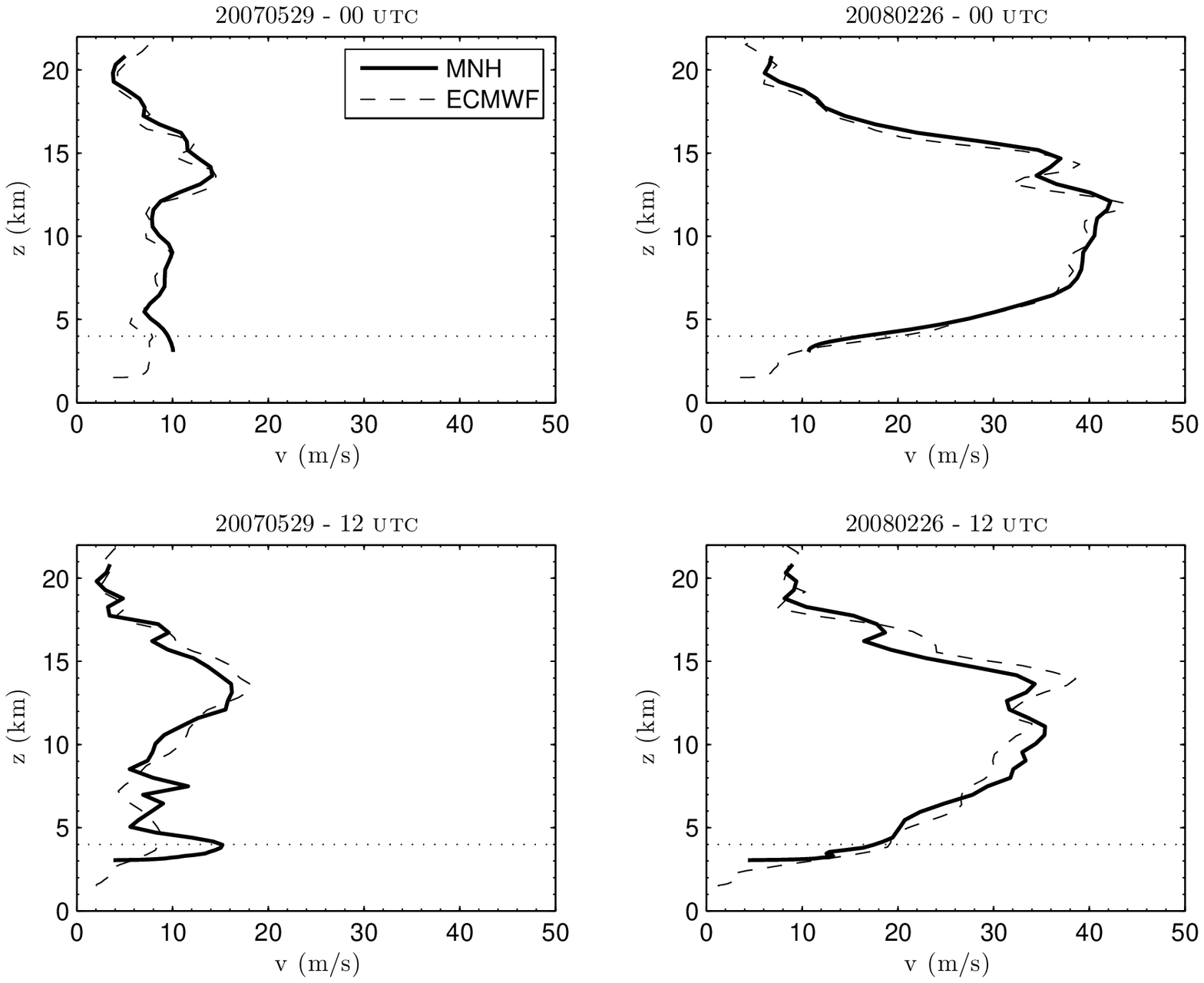}
\caption{Vertical wind speed profile of the Meso-NH (solid line) and
  the ECMWF analyses (dashed line) at the beginning (00:00 UTC) and
  the end (12:00 UTC) of a simulation for two nights, May 29, 2007
  (left hand side) and Feb. 26, 2008 (right hand side).}
\label{wind_ws}
\end{figure*}

What is the Meso-NH ability in reconstructing the temporal evolution
of the wind speed? To answer to this question we choose two nights,
one characterized by a weak wind speed (May 29, 2007) and another
characterized by a strong wind speed (February 26, 2008), and we
compared the wind speed profiles calculated with the ECMWF analyses
and the Meso-Nh model at the start and at the end of the
simulation. We consider the wind speed calculated by the Meso-NH at
the peak of Mt.Graham and the ECMWF wind speed extracted in the
nearest grid point to the Mt. Graham summit. Again we are only
interested, in this phase, in the wind speed above 1 km from the
ground and, following the same logic we used in Section \ref{ses2}, we
discuss the results obtained above 4 km from the sea
level. Figure~\ref{wind_ws} shows the result of the comparison. ECMWF
profiles are interpolated to the Meso-NH vertical grid
points. Fig.~\ref{wind_ws}-top and Fig.~\ref{wind_ws}-bottom show,
respectively, the start and the end of the simulation We observe that
the wind speed profile calculated by Meso-Nh has evolved during both
of the nights accordingly with the wind speed as calculated by the
ECMWF. The shape of the profile is reasonably well correlated as well
as the strength of the wind speed. We observe that the wind speed
increases somewhat during the night on May 29, 2005 while decreases
its strength on February 26, 2008.  We conclude therefore that the
mesoscale model is able to reconstruct the wind speed vertical
distribution in the high part of the atmosphere and reproduces the
spatio-temporal wind speed variability in a satisfactory way. In the
case of May 29, 2007 (Fig.~\ref{wind_ws}-bottom) the wind speed in the
low part of the atmosphere reconstructed by the Meso-Nh model presents
a few peaks that might be originated by the better horizontal
resolution of the mesoscale model.


What about the ability of Meso-Nh in reconstructing the wind speed in
the first kilometer from the ground?  For this study we consider
measurements provided by two different instruments: (1) measurements
of the wind speed profile made with a Generalized Scidar \citep{Ma10}
and related to 39 nights in different seasons (Table \ref{tab2}).  (2)
'in situ' measurements of the wind speed done with an anemometer located
on the roof of the VATT ($\sim$ 20 m from the ground). 

For these nights we can retrieve from the GS the wind speed along the
whole 20 km. As already mentioned, a comparison of the Generalized
Scidar wind speed profiles and the ECMWF analyses at Mt, Graham has
been done in \cite{Eg07} and it has been found a good correlation for h
$\ge$ 1 km.  Qualitative comparisons of the Generalized Scidar wind
profiles with NCEP/NCAR re-analysis and wind speed from in situ
balloons for 15 nights in May 2000 at San Pedro M\'artir has been
described in \cite{Av06}. Also \cite{GL06} compared the Generalized
Scidar wind speed measurements done at the Teide Observatory with
radiosoundings data launched by a meteorological station placed at
$\sim$ 13 km away from the summit for four summer nights in
2003. However, to our knowledge, there is no detailed study of the
reliability of the Generalized Scidar at describing the wind speed in
the surface or boundary layer. For this reason, the measurements from
an anemometer have been used to have an independent measurements in
the very low atmosphere.

\begin{table}
\caption{Observing runs at Mt.Graham and the number of nights for
 which there are wind speed profiles from the GS.}
\begin{tabular}{cc}
\hline
Obs. runs ({\sc utc}) & Nights \\
\hline
27 April 2005 & 1\\
20-26 May 2005 & 6 \\
7-16 December 2005 & 5 \\
28 May - 4 June 2007  & 8 \\
17-29 October 2007 & 11 \\
24 February - 4 March 2008 & 8 \\
\hline
\label{tab2}
\end{tabular}
\end{table}

\begin{figure*}
\includegraphics[width=7cm]{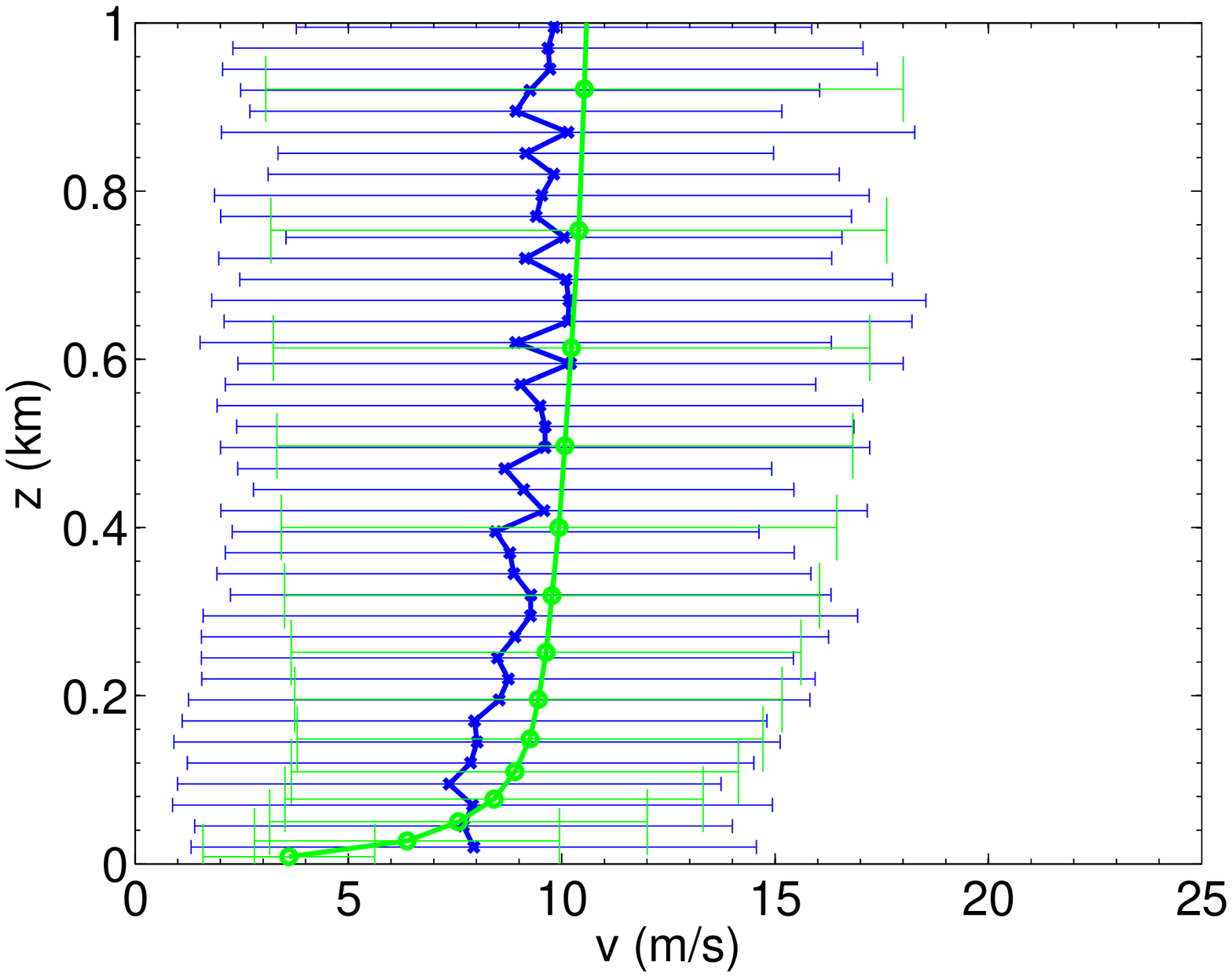}
\includegraphics[width=7cm]{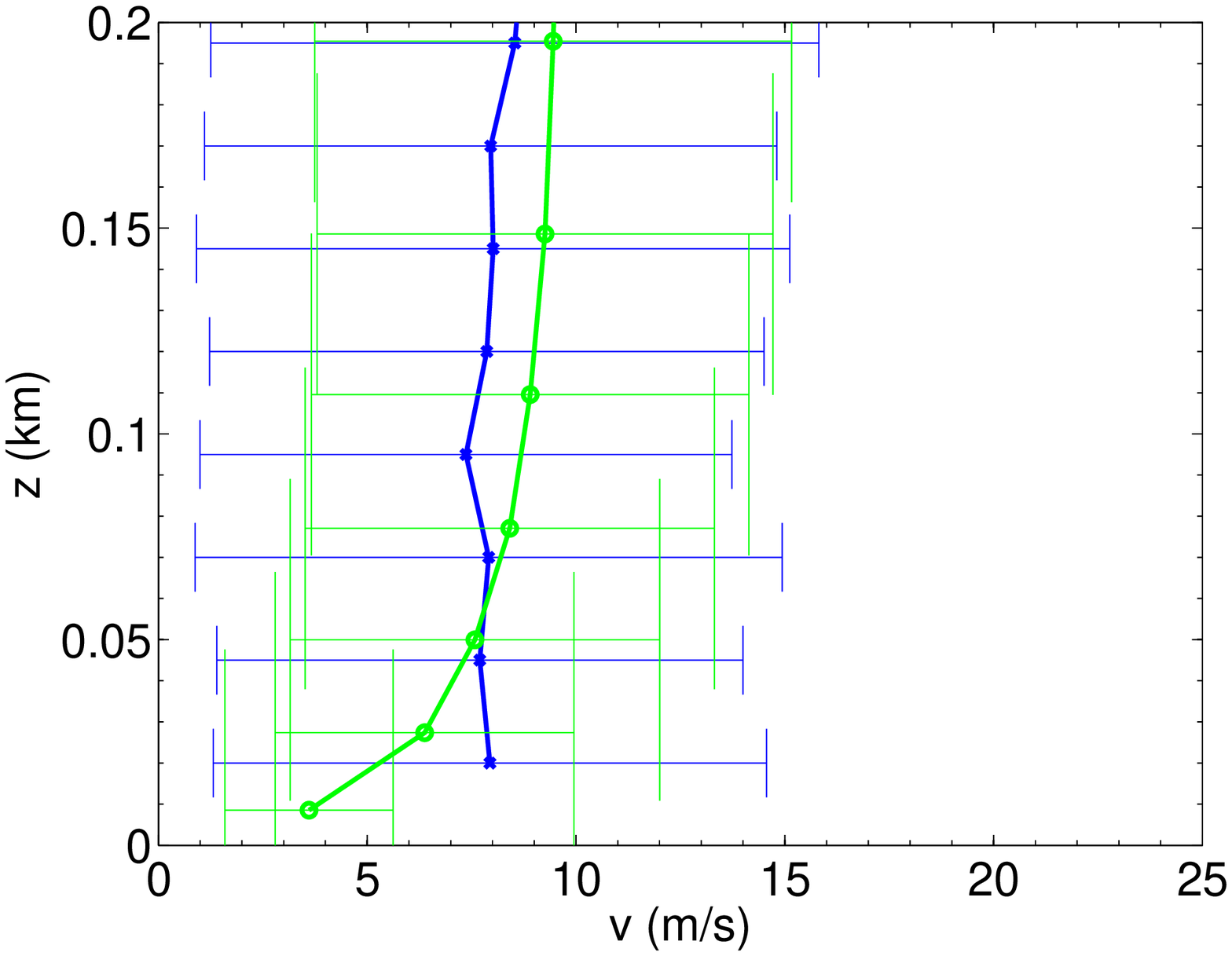}
\caption{Left: Mean vertical profile of the wind speed in the boundary layer (1 km)
 obtained with the GS (blue line) and the Meso-NH (green line). The
 error-bars show the standard deviation. Right: zoom in the first 200 m.}
\label{wind_gs}
\end{figure*}

\begin{figure}
\includegraphics[width=7.5cm]{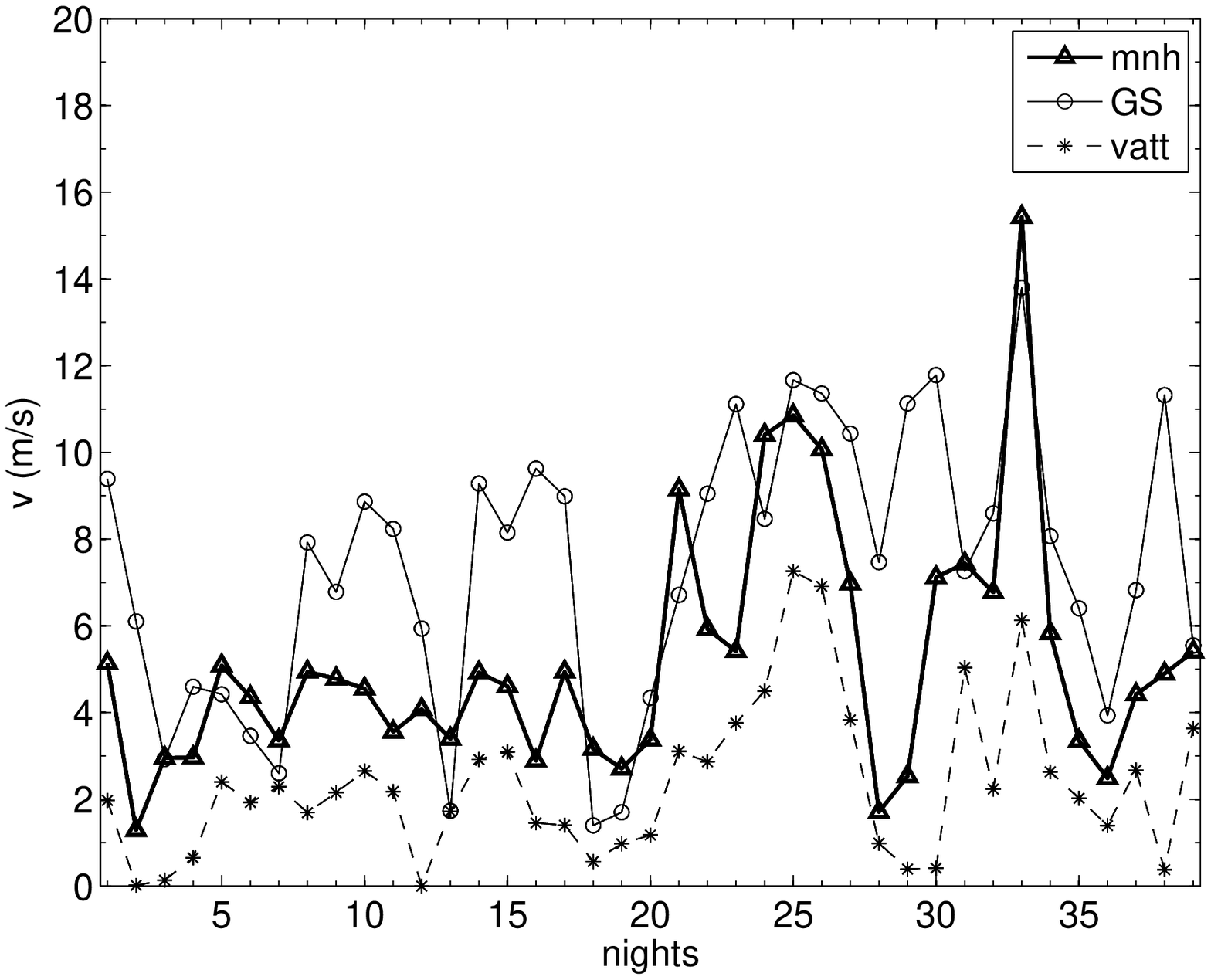}
\caption{The average wind speed near the surface for Meso-NH (thick
 solid line and triangles) at h=20 m, an anemometer placed on the roof of the VATT at h $\sim$ 20m (dashed line
 and stars) and the Generalized Scidar (thin solid line and circles).}
\label{wind_surf}
\end{figure}

We have calculated the average wind speed vertical profile in the
first kilometer for all 39 nights for which there are measurements of
the wind speed from the Generalized Scidar (Fig.~\ref{wind_gs} - blue
line) and compare the resulting profile with the average wind speed
simulated by the Meso-NH for the same nights (Fig.~\ref{wind_gs} -
green line). We remind that the GS is placed at the focus of the
VATT\footnote{The Vatican Advanced Technology Telescope is also
  located at the top of Mt. Graham at around 250 m from the LBT.}
therefore the first measurement (on the vertical grid) for the GS is at $\sim$ 20 m
from the ground.  
Looking at Fig.\ref{wind_gs} we observe that, almost everywhere the two
profiles are very well correlated, with the Generalized Scidar
estimating wind speed that differs from the Meso-Nh model for not more
than 1 ms$^{-1}$ all along the first kilometer from the ground. At 20 m the GS wind is slightly stronger than the wind calculated by the model.
This is probably due to the fact that the GS measurements have a vertical error bar of
the order of 25-30 m (see Masciadri et al. 2010) and the wind shear is particularly strong at this height.
Therefore it might be
that the higher wind speed detected by the GS at h =20 m can be
originated by thin layers flowing at slightly higher distance from the
dome and not resolved by the instrument. Figure \ref{wind_surf} shows the average wind speed
near the ground for every night reported in Table \ref{tab2} in
chronological order as measured by the Generalized Scidar and
reconstructed by Meso-NH at around 20 m i.e. at the height of the dome of the
telescope. In the same figure is also reported the wind speed measured
in the same nights by the anemometer mounted on the top of the VATT. 
From a qualitative point of view, in Fig.~\ref{wind_surf} we
observe that the temporal evolution of the wind speed reconstructed by
the model during the 39 nights is very well correlated to the
anemometer trend. The wind speed reconstructed by the model decreases
and increases following the wind speed evolution measured by the
anemometer in the same nights. Only a small off-set is present between
the two estimates. From a quantitative point of view the mean wind speed from the GS at h = 20 m is
7.37 ms$^{-1}$, the mean wind speed reconstructed by the Meso-Nh model
is 5.21 ms$^{-1}$, the mean wind speed measured by the anemometer is
2.44 ms$^{-1}$.  The wind speed reconstructed by the model is
well included in the range of the wind speed
measurements and this certainly proves the reliability of the calculated 
wind speed. The weaker wind speed from the anemometer
is probably due to the fact that the anemometer is an 'in situ' measurement
that is done well below the top of the trees. 
It has been observed in the past that the friction of the atmospheric flow 
with the trees causes a sharp and drastic decreasing of the wind speed below this height (something that is 
confirmed in the profile reconstructed by the model below 20 m).

We conclude therefore that the model provides
reliable estimates all along the whole 20 km. 
In the future it would be good to be able to equip the observatory with anemometers 
located at different heights below and
above the top of the trees preferably in open space environment so
that the wind speed is not affected by the presence of buildings. This should permit us to better 
monitor the particular sharp change of wind speed in this region. 

\section{Conclusions}


In this paper we aimed to give a complete characterization of the
wind speed vertical distribution at Mt. Graham (Arizona, US) for astronomical 
applications. The simplest way of retrieving a complete characterization of the 
wind speed profile (with exception of the surface and boundary layer) 
is using the data from General Circulation Models (GCMs). In this study we use
the operational analyses from the ECMWF extracted from the grid point
closest to the peak of Mt. Graham, with a 0.25 degrees resolution, to
study the vertical wind speed distribution over 10 years (1998-2007).

We have verified that the wind speed profile retrieved from the
operational analyses of the ECMWF model is consistent with what is
obtained by the radiosoundings from the nearby Tucson International
airport. We also proved that the wind speed in this region is
homogeneous with respect to horizontal spatial scales of the order of
some tens of kilometers above $\sim$4 km from the sea level.

Having proved that the ECMWF analyses are reliable, we presented the
monthly median wind speed extended on a 10 years time scale. The wind
speed profiles are rather typical of a mid-latitude site with a
pronounced wind speed maximum at the jet stream level (10-12 km above
sea level) during most of the year. On the contrary, during the summer
the maximum wind speed is located well into the stratosphere. The
strongest variability (from different years) in the monthly median
wind speed is found during spring and autumn.

For the same period we also provided the monthly mean wind speed
values at 200 hPa, corresponding to the maximum wind speed values at
the jet-stream height. The month with the strongest wind speed at 200
hPa is February (37.21 ms$^{-1}$), while the weakest wind speed is observed
in July (11.17 ms$^{-1}$). Results indicate that the wind speed at the
jet-stream level is very similar to what has been observed
above the Observatory in San Pedro M\'artir (Baja California)
and it is consistent with the values observed above the best
astronomical sites in the world. 

Besides, we proved the reliability of a mesoscale model (Meso-Nh) in
reconstructing the wind speed on the whole 20 km just above the summit
of Mt. Graham included the boundary layer and the surface layer. We
proved that the wind speed reconstructed by the model is very well
correlated to the ECMWF analyses and it is also
able to provide realistic profiles in the first kilometer from the
ground.

The wind speed estimates from the mesoscale model have been compared
to measurements from a Generalized Scidar and an anemometer located at
20 m from the ground on a sample of 39 nights.  Above 50 m the wind
speed profiles reconstructed by the model match in a very satisfactory
way ($\Delta$V $\le$ 1 ms$^{-1}$) with respect to the measured wind
speed profiles.  Closer to the surface, just in proximity of the top
of the trees, the wind speed estimated by the model is included in the
range of values given by the anemometer and the Generalized Scidar and
for this reason can be considered satisfactory. However, the
dispersion between the anemometer and the GS measurements seems a
little too large. This difference is highly probably due to the fact the anemometer
measures a wind speed that is not completely in free air (the
anemometer is placed beside the dome of the VATT) and below
the top of the trees. On the other side, the GS is probably affected
by the wind speed just above the top of the trees because of its
finite vertical resolution. The qualitative trend of the wind speed
observed all along the sample of 39 nights is however very well reconstructed
by the model and in agreement with measurements. The model appears to
reconstruct very well the wind speed behavior above and below 20 m.

This paper therefore validates the Meso-Nh model as a tool to predict
the wind speed vertical profile V(h) from the ground up to 20 km above
Mt. Graham for each night and, at present time, it appears as the
unique method to systematically estimate the whole wind speed vertical
profile above an astronomical Observatory. We remind that it has been proved 
(Masciadri et al. 1999a, Masciadri \& Jabouille, 2001, Masciadri et al. 2004, Lascaux et al. 2010) that Meso-Nh 
can provide reliable $\CN2$ profiles above an astronomical site and it appears therefore 
as an extremely useful tool for $\tau_{0}$ estimates. Besides there are evidences
that it would be very useful to supply the Mt. Graham Observatory with
anemometers located at different heights below and above the top of
the trees ($\sim$ 20 m) because this should permit to provide a better
constraints of the model for dedicated and more detailed applications.

\section*{Acknowledgements}
ECMWF products are extracted from the catalogue MARS,
http://www.ecmwf.int, access to these data was authorized by the
Meteorologic Service of the Italian Air Force. Radiosoundings are
obtained from the University of Wyoming-site
http://weather.uwyo.edu/upperair/sounding.html. This study has been
funded by the Marie Curie Excellence Grant (FOROT) -
MEXT-CT-2005-023878.

\appendix

\section{The wind speed at different horizontal resolution}

\begin{figure*}
\includegraphics[angle=-90,width=175mm]{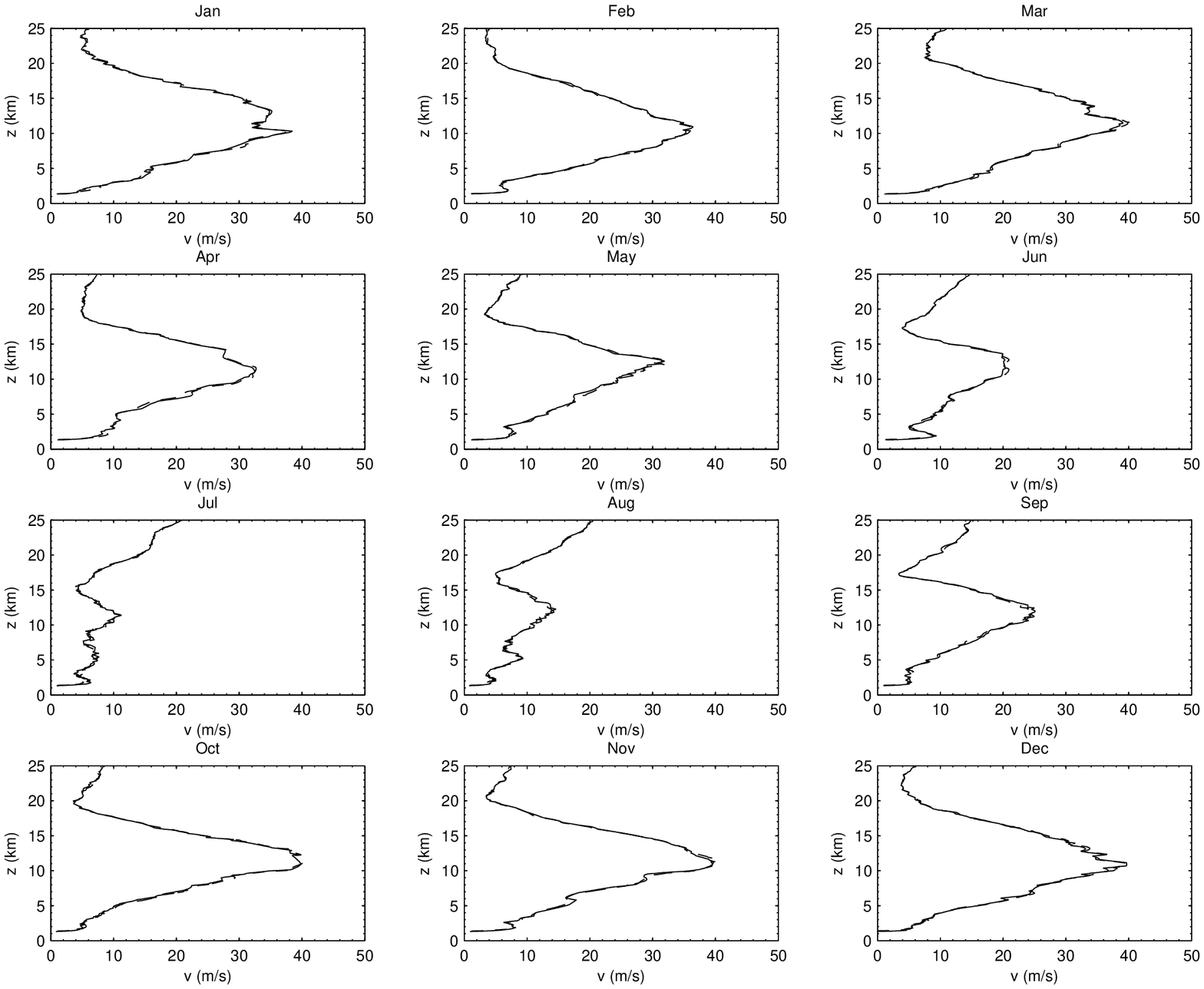}
\caption{The median monthly wind speed for 2002 at the two different grid
 points closest to the peak of the mountain using 0.25$\degr$
 resolution (32.75$\degr$N, 110.00$\degr$W) (solid lines) and
 0.5$\degr$ resolution (32.50$\degr$N, 110.00$\degr$W) (dashed lines). 
\label{wind2002}
}
\end{figure*}

The horizontal resolution of the operational analyses of the ECMWF has
changed during the ten year-period studied in this paper. This implies
that the position of the closest grid points to Mt. Graham
changes. The data in this study are downloaded from the 0.25$\degr$
resolution where the closest grid point to Mt. Graham is located east
northeast of the mountain peak (Fig.~\ref{map} - point (A)) at
32.75$\degr$N, 110.00$\degr$W. Using the 0.5$\degr$ resolution the
closest grid point is located 32.50$\degr$N, 110.00$\degr$W, southwest
of the mountain (Fig.~\ref{map} - point (B)).

To closer examine which impact the different resolutions have on the
vertical wind speed profile we have downloaded the data from both of
these grid points for the entire year 2002. The monthly median wind
speed profile is presented in Fig.~\ref{wind2002}, where the data from
the grid point using the higher resolution (32.75$\degr$N,
110.00$\degr$W) is plotted with a solid line and the data from
the 0.5$\degr$ resolution (32.50$\degr$N, 110.00$\degr$W) is plotted
using a dashed line. The difference between the two data-sets is very
small. During all months the lines overlap each other almost
entirely. Some smaller offsets exist, but they are generally
minor. The largest difference found, near the jet stream-level in
October, is 2.8 ms$^{-1}$. We conclude therefore that the change in
horizontal resolution in ECMWF-analyses during the 10 years did not
introduce any biases in our calculation.

\label{lastpage}
\end{document}